# Diffuse Optical Ptychography

Mingwei He[1], Sujit K. Sahoo[2*], Chengyuan Xiao[1] and Cuong Dang[1*]

[1] School of Electrical and Electronic Engineering, Nanyang Technological University Singapore, 50 Nanyang Avenue, 639798, Singapore

[2] School of Electrical Sciences, Indian Institute of Technology Goa, Goa 403401, India

*Corresponding author, Email: hcdang@ntu.edu.sg, sujit@iitgoa.ac.in

## Abstract

Various imaging techniques have significantly enhanced our ability to visualize objects embedded within complex media such as biological tissues, fog, atmosphere, or various turbid media. Optical imaging, in particular, offers multiple advantages, including non-invasive capabilities, absence of ionizing radiation, and high contrast for many biological tissues. However, optical imaging through substantially thick scattering media remains challenging due to extensive photon diffusion, significantly restricting reconstruction quality and achievable resolution. To address these limitations, we introduce Diffuse Optical Ptychography (DOP), a novel imaging method inspired by ptychography technique, which exploits additional spatial information gained from multiple overlapping illumination patterns. The primary technical innovation of DOP lies in its effective use of overlapping yet minimally correlated illuminations, significantly enhancing reconstruction accuracy and image quality. Compared to existing optical imaging methods through thick diffusive media, DOP achieves superior resolution (down to 1 mm) and reliably reconstructs both binary and grayscale objects embedded within media thicker than 100 transport mean free paths. Importantly, DOP demonstrates robust reconstruction performance both with accurately calibrated diffusion properties and even without prior calibration. Furthermore, the experimental setup for DOP remains straightforward, utilizing only a conventional camera and scanning illumination spots. Our demonstrations underscore the broad potential impact of DOP in applications ranging from medical diagnostics to non-destructive testing, thus opening promising avenues for high-resolution imaging in highly scattering environments.

## Introduction

Conventional optical imaging methodologies, such as photography or microscopy, capture clear object images directly with a sensor array. However, scattering media with complex refractive index distributions and reflectance properties complicate the imaging process by scrambling optical paths[1]. Scattering media, such as biological tissues, fog, and turbid liquids, are abundant in nature and create substantial obstacles across many disciplines, notably in medical diagnostics or non-destructive testing applications. Developing effective imaging techniques to overcome these scattering-induced challenges will have significant impacts.

Photon propagation through scattering media inherently involves random walks within a three-dimensional space, posing considerable difficulties for imaging as transmission depth increases. For relatively thin scattering media, where photons typically experience single or limited scattering events, advanced techniques such as wavefront shaping[2-7], speckle imaging[8-12], and machine learning-enhanced approaches[13-16] have been developed, demonstrating superior and even super-resolution imaging capabilities[17-18]. In thicker scattering media, imaging complexity substantially escalates. Techniques like time-gated imaging[19-21], polarization filtering[22-23], or scattering event tracking[24-25] attempt to extract information from the photons that experience few or no scattering events, referred to as snake or ballistic photons. However, when the thickness of scattering media significantly exceeds the transport mean free path, $l_t$, the medium becomes highly diffusive[26-27]. The transport mean free path represents the average distance photons travel before losing directional memory due to successive scattering events[28], hence, there are no snake photons or ballistic photons in diffusive media (Fig. 1a). In biological tissues, typical absorption and reduced scattering coefficients are approximately $\mu_a = 0.05\ cm^{-1}$ and $\mu'_s = 10\ cm^{-1}$, respectively, yielding the transport mean free path around $l_t = \frac{1}{\mu_a + \mu'_s} = 0.1\ cm$[29]. Consequently, imaging through diffusive biological tissues, even with a thickness of just a few millimeters to a centimetre, becomes highly challenging.

Diffuse Optical Tomography (DOT) techniques have shown potential by reconstructing images through highly diffusive media; however, resolution remains limited, typically for centimetre-scaled objects buried under several centimetres of tissue thickness[30]. Recent demonstration, a time-of-flight DOT approach (ToF-DOT), achieved millimeter-level resolution for simple binary objects embedded within media as thick as 50mm ($l_t$ = 600 μm)[31]. This method utilizes ultrafast lasers and single-photon avalanche diode (SPAD) arrays to capture both spatial and temporal photon distribution. Nevertheless, practical adoption remains limited due to high equipment costs, the complexity of ultrafast laser systems, SPAD cameras' low resolution and elevated noise. Furthermore, this technique heavily depends on prior knowledge of scattering properties for accurate point spread function (PSF) estimation, creating additional implementation barriers.

Here, we introduce Diffuse Optical Ptychography (DOP), a novel method designed to image objects deeply embedded within highly diffusive media. Our approach offers two main advancements over existing techniques. First, we eliminate the complexity and expense of ToF-DOT imaging by demonstrating that temporal information contributes minimally or not at all to imaging in highly diffusive conditions. Second, DOP algorithm utilizing extra spatial information, while remaining a simple experimental setup (Fig. 1a-b), reliably recovers embedded objects with or even without prior scattering knowledge as presented in Fig. 1e, respectively. Inspired by traditional ptychography, which uses sequential illumination of overlapping sample regions to boost reconstruction accuracy[32-33], our DOP leverages extra spatial information (Fig. 1c and 1d) from multiple overlapping illuminations. Our DOP successfully reconstructs binary and grayscale objects obscured by scattering media thicker than 100 transport mean free paths (approximately 32 mm, while $l_t$ = 300 μm), achieving a remarkable resolution of 1 millimeter.

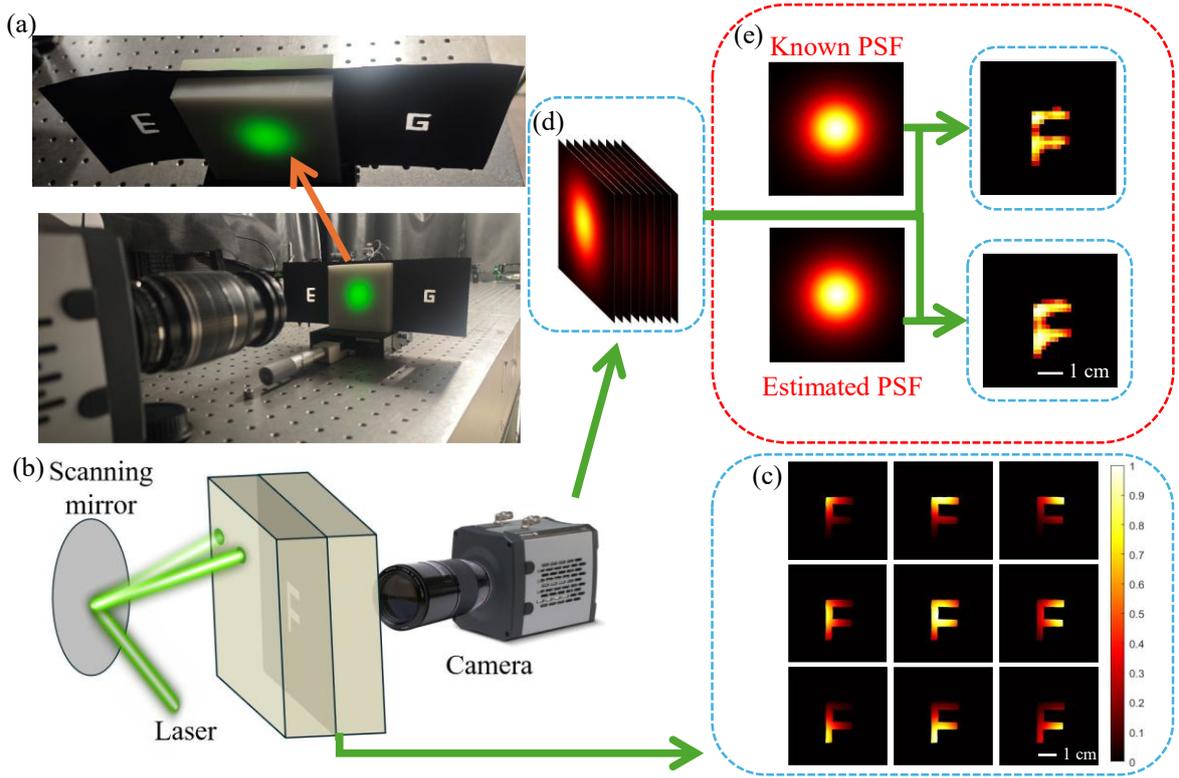

**Figure 1. Experimental setup and workflow diagram. a)** Experimental setup on the lab's optical table where the object is sandwiched between 2 diffusive blocks, with the thickness of 32 mm (~100 $l_t$). **b)** Schematics of DOP setup where a scanning mirror guides a laser beam to desired positions on the surface of the first diffusive block; the multiple illumination points are within a 3x3 cm square. A CMOS camera is placed on the opposite side of the laser to collect the diffused transmission light. **c)** A series of images illustrating light transmission through the object, each corresponding to a different laser illumination point. **d)** A series of captured images, one for each laser illumination point. **e)** DOP algorithm reconstructed the hidden object with or without diffusion calibration. In the absence of diffusion calibration, PSFs are initially approximated as Gaussian patterns and subsequently refined using diffusion solutions.

# Time-integrated Diffuse Optical Tomography

We first revisit the ToF-DOT to evaluate the contribution of temporal information in image reconstruction. The SPAD camera measures the 3D spatial-temporal data cube of photon distribution after impulse illumination by an ultrafast laser. Mathematically, with noise-free measurements, if each temporal frame presents an independent measurement of the same object, such 3D dataset will significantly enhance reconstruction quality. Conversely, if temporal frames are completely correlated, redundant information will not improve image reconstruction. Evaluating the temporal information contribution to imaging is complex due to the strong relation between spatial and temporal distribution in the diffusion process.

For simple evaluation, we utilize the experimental 3D data cube from previous ToF-DOT studies[31]. If we replaced the SPAD camera in the ToF-DOT setup with a conventional camera of identical spatial resolution, we would capture a single time-integrated image (2D data), representing the summed temporal data from the SPAD camera (Fig. 2a), i.e. removing all the temporal information of the data cube. Figure 2b shows the reconstruction results using this single time-integrated image, for which we call DOT to differentiate from ToF-DOT reconstruction results in Fig. 2c. It is worth noting that both reconstructions were obtained by directly running the provided algorithm[31] with the respective 2D and 3D datasets, without any adjustments or contrast enhancements (the contrast-enhanced images are available in Supplementary Fig. S1). Certainly, the computational cost for running with the 2D dataset is less than that of the 3D dataset by 2 orders.

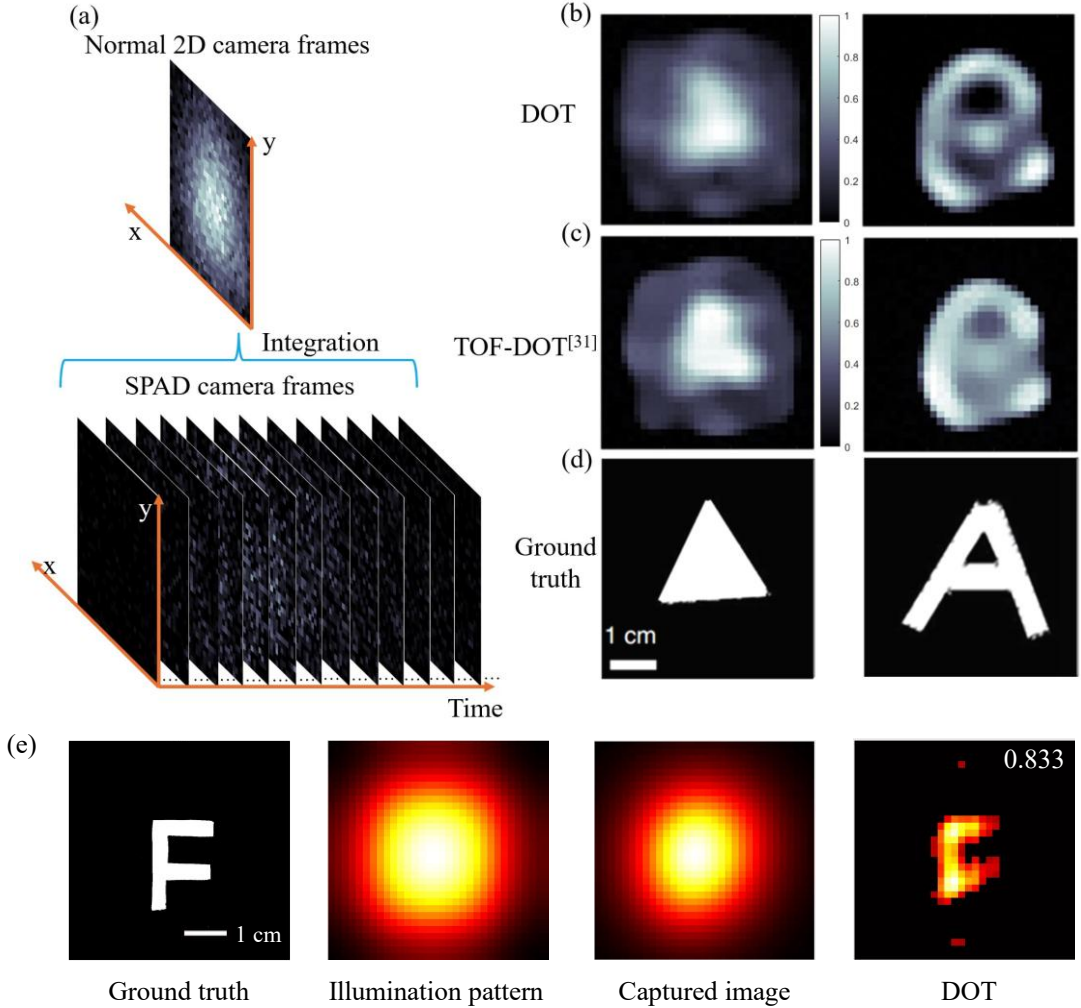

**Figure 2. Experimental results with time-integrated measurements and time-resolved measurements. a)** The summation of all temporal frames from the SPAD camera produces a single 2D image, equivalent to a conventional camera image. **b)** The DOT result with a single 2D image as input for reconstruction. **c)** ToF-DOT result reconstructed with all temporal frames. **d)** Ground truth. Figure b, c, d share the same scale bar as presented in figure d. **e)** The DOT result obtained using our experimental setup with a conventional camera, along with its correlation coefficient relative to the ground truth; they all have the same scale bar as presented in the ground truth image. Illumination pattern is the laser illumination pattern on the front surface of the first diffusion block.

Interestingly, the conventional camera approach yields better results compared to ToF-DOT (Fig. 2b-d). Despite originating from the same data source, the integrated image has a higher signal-to-noise ratio (SNR) due to the summation of multiple temporal frames, and no timing jitter issues associated with SPAD cameras. Our simulation results in supplementary Fig. S2 show equivalent image quality for conventional DOT and ToF-DOT, reinforcing the conclusion that temporal information in ToF-DOT is negligible and can even degrade imaging quality due to the associated hardware's noise and jitter. This result is understandable if we look into the correlation among the temporal frames. We simulate the photon diffusion process to obtain spatiotemporal information under the same experimental conditions to eliminate the impact of SPAD camera's limitation. When analyzing all temporal frames that account for 99% of the total spatiotemporal signal, the correlation coefficient among frames consistently exceeds 94.5% (see Supplementary Fig. S3). Similarly, high correlations are also obtained for the temporal diffusion patterns at the embedded sample (Supplementary Fig. S4). These findings imply that despite the impulse laser beam shining from the front, we still illuminate the sample with highly correlated temporal patterns and obtain highly correlated temporal frames at the output due to the photon's diffusion nature. Thus, temporal information provides minimal additional value in this extremely high diffusion scenario.

We perform DOT with a conventional camera in our setup (Fig. 1 and Supplementary Note 1), where the laser illuminates the sample from the first diffusive layers. Fig. 2e presents the results serving as benchmarks for evaluating our subsequent imaging results.

## Diffuse Optical Ptychography (DOP)

Traditional ptychography uses a coherent beam to sequentially scan the sample with overlapping illuminated regions[32-33]. Another version of it, Fourier-ptychography does not require a coherent beam to scan the sample's Fourier plane with overlapping illuminated regions[34-35]. After striking the sample, the beam propagates through a free space or an imaging system to form patterns on the detector. The overlapping regions produce extra information to enhance the reconstruction quality compared to a single measurement with a single illumination. Inspired by this, we propose DOP, which utilizes extra information from overlapping illuminated regions to gain reconstruction quality. Unlike typical ptychography, we do not have direct access to the sample from both the illumination and the observation sides, instead, the light is diffused to and from the sample through thick scattering blocks. Unlike ToF-DOT, all multiple measurements in our DOP contribute comparable intensity with significantly lower correlation (Supplementary Fig. S5), expecting to provide significant extra information to enhance reconstruction quality.

A uniform diffusive medium is characterized by a shift-invariant point spreading function (PSF). Incident illumination, $I_i$, propagates to the object $O$ through the first diffusive block (characterised by $PSF_1$). By varying $I_i$, the object's illuminated region, calculated as $(I_i * PSF_1) \odot O$ (where $*$ is convolution and $\odot$ is the element-wise multiplication), is expected to be both overlapped and uncorrelated with each other. The illuminated light continues propagation through the second diffusive block, characterized by $PSF_2$, before being captured by a camera. The measured signal is $M_i = [(I_i * PSF_1) \odot O] * PSF_2 + W$, where $W$ represents noise. DOP algorithm solves this inverse problem to reconstruct $O'$ as close as possible to the object $O$ by minimizing the following expression.

$$\underset{O', PSF_1, PSF_2}{\text{Minimize}} \frac{1}{2}\Sigma_i ||[(I_i * PSF_1) \odot O'] * PSF_2 - M_i||_2^2 + C(O') \quad (1)$$

where $C(O')$ is the constraint function. Here, we use sparsity and continuity constraint functions for our object (more details in Supplementary Note 2 and Fig. S8). Each PSF is characterised by diffusion properties ($\mu_a, \mu_s'$) and the diffusion thickness ($d$), which might be known or unknown depending on the actual scenario or application. Biological tissue might be difficult to characterize; therefore, it is ideal to do imaging with many unknowns. However, for non-destructive inspection applications where the defects are embedded inside known materials, the diffusion characteristics can be well-determined.

**DOP with diffusion calibration.**

In this section, we discuss DOP in the simplest condition, i.e. all 6 diffusion parameters of the two diffusive blocks are pre-calibrated, therefore, both PSFs are computed precisely. Because the key requirement for DOP is the overlap and uncorrelation of the object's illuminated region $((I_i * PSF_1) \odot O')$, we choose point illumination patterns, $I_i$, by scanning the laser beam at different positions on the first block (Supplementary Note 1). Figure 3 presents four different sets of illumination patterns and their corresponding DOP results. The object is the letter F, similar to Fig. 2e. Within the same illumination area, the number of illumination points ($I_i$) increases from 4 to 9, i.e. the overlapping regions increase progressively, leading to enhancement of reconstruction quality. The results highlight the roles of overlapping regions in reconstruction quality, the signature of ptychography. Figure 3d and Supplementary Fig. S6, showing superior reconstruction quality, present the key achievement of our DOP. Increasing the number of illumination points beyond 3x3 for DOP does not enhance the reconstruction quality further because the illumination patterns on objects vary very little among the adjacent illumination points, i.e. highly correlated illumination patterns on the object. This is further illustrated in the correlation matrix of all the measured patterns at the output plane (Supplementary Fig. S5). The correlation is very high among the captured images with the adjacent illumination points. Compared to DOT with a single measurement (Fig. 2e), DOP is superior. It is worth noting that the single measurement in DOT is done with different illumination strategies such as an enlarged Gaussian laser beam, or a long exposure imaging time during which the laser scans all 9 points, or a summation of all 9 measurements in Fig. 3d. All the reconstruction qualities are similar to Fig 2e. However, the single measurement for DOT in Fig. 4 below are the summation of all the 9 measurements corresponding to 9 illumination points so that both DOT and DOP have the same data source for fair comparison.

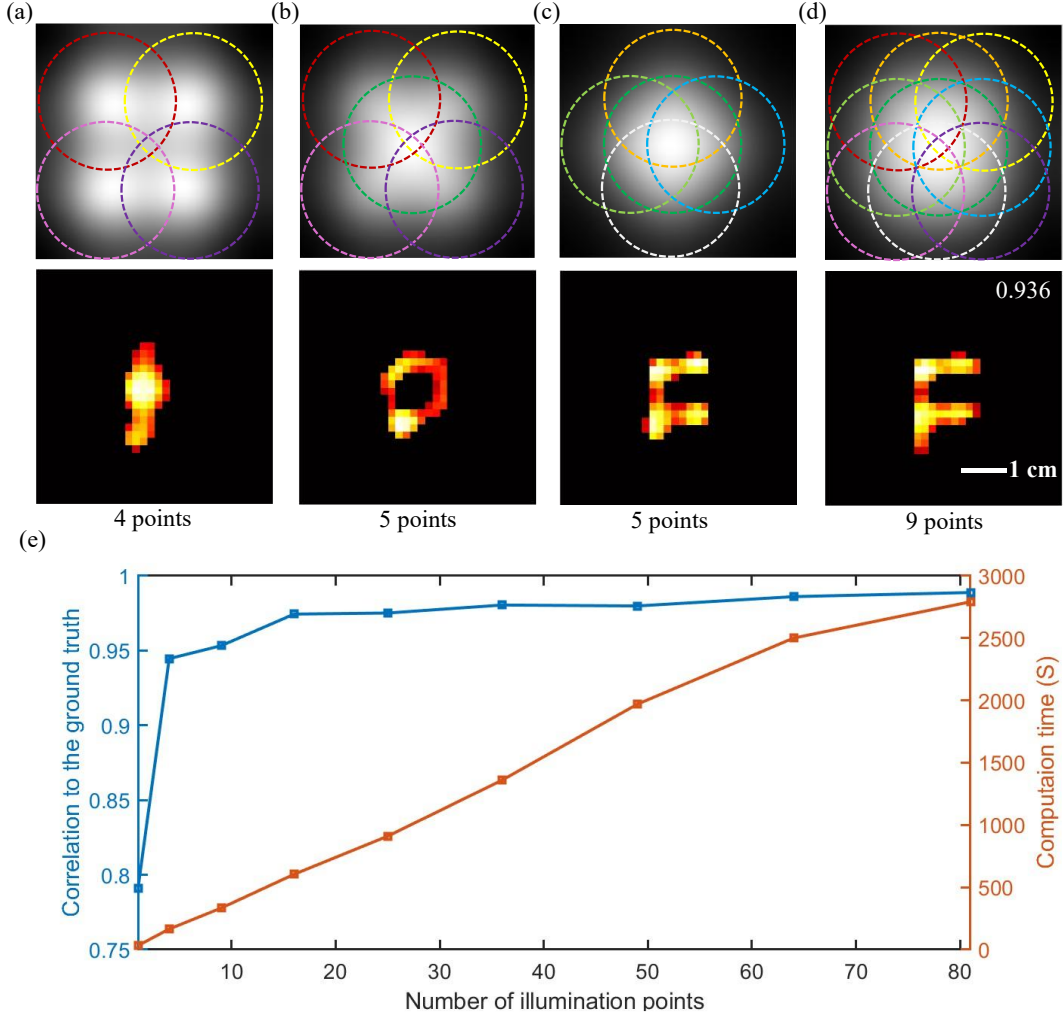

**Figure 3. DOP results with different illuminations when diffusion characteristics are known. a-d)** DOP with the illumination points increases from 4 to 9. The top row presents the illumination patterns at the object plane. The object and the second diffusive block are removed to take these images for illustration purposes only (DOP algorithm does not use these images). The dash circles present individual illumination patterns, once at a time in DOP process. The bottom row shows the DOP results corresponding to the illumination patterns in the top row. The number in figure d indicates its correlation coefficient relative to the ground truth. **e)** DOP reconstruction quality and computational cost as functions of the illumination point number.

**DOP with unknown diffusion characteristics.**

In many practical scenarios, the diffusive media are often either unknown or characterized with limited precision, leading to inaccuracies in the estimated PSFs and challenging the reconstruction algorithm. In this case, we need to solve the optimization (1) to find not only $O'$ but also diffusion characteristics for both PSFs. Many previous studies approximated the PSF as a Gaussian pattern[36-37], but the fine difference between them limits the construction quality and resolution is typically limited to cm for a few cm diffusion layers. Here, the extra information from DOP could gain better PSF and $O'$ estimation. First, we simplify both PSF$_1$ and PSF$_2$ as Gaussian patterns, allowing us to reduce the optimization space for PSFs to 2 dimensions ($\sigma_1$, $\sigma_2$) from 6 dimensions ($\mu_{a1}, \mu'_{s1}, d_1, \mu_{a2}, \mu'_{s2}, d_2$). We run an individual optimization process for each pair of σ$_1$ and σ$_2$, which are scanned on a large scale to cover all the PSF possibilities, and then the final result will be global optimization for $O'$, $\sigma_1$ and $\sigma_2$. The input for the optimization algorithm is either a single-shot image or multiple illumination images, whose outputs are presented in Fig. 4b and 4c, respectively. The results highlight the significant impact of multiple overlapping illuminations in reconstruction quality – the signature of Ptychography. However, the quality is still not very high due to the Gaussian approximation of PSFs.

From optimized Gaussian PSFs, we estimate diffusion PSFs with the diffusion solution[38-39] (Supplementary Eq. S1, S2). We select narrow ranges of $\sigma_1$, $\sigma_2$ around the optimized values of Gaussian PSFs. Then we estimate the set of 6 parameters ($\mu_{a1}$, $\mu'_{s1}$, $d_1$, $\mu_{a2}$, $\mu'_{s2}$, $d_2$) so that the resulting diffusion PSFs are best fit with Gaussian PSFs characterized by a pair of $\sigma_1$, $\sigma_2$. Note that we can easily measure the total thickness $d$, then calculate $d_2 = d-d_1$ to reduce the optimization parameter from 6 to 5. From each set of 5 parameters, we run gradient-descent optimization to find locally optimized $O'$, the final result of $O'$ is the global optimization for all parameter sets. The input for the optimization algorithm is either a single-shot image or multiple illumination images, whose outputs are presented in Fig. 4d and 4e, respectively. Again, our DOP provides significant improvement in reconstruction quality compared to typical DOT. Comparing Fig. 4c with Fig. 4e, it is obvious that a better estimation of PSFs from diffusion solutions enhances the image quality significantly. More results for different diffusion characteristics are presented in Supplementary Fig. S7. The demonstrations show the superior reconstruction quality of our DOP, where no diffusion properties are required to be pre-measured.

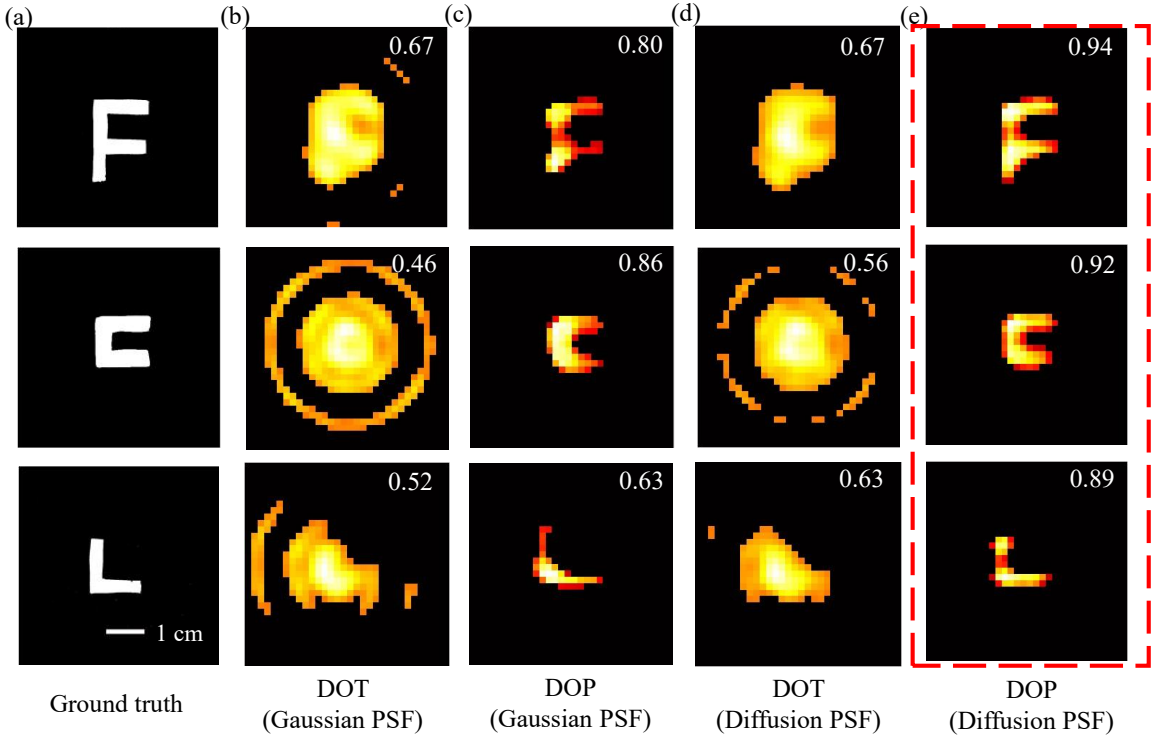

**Figure 4. DOT and DOP reconstruction results without diffusion calibration. a)** Ground truth. **b)** DOT and **c)** DOP reconstruction with Gaussian approximated PSFs. **d)** DOT and **e)** DOP reconstruction with estimated diffusion PSFs. All reconstructed images share the same scale bar as shown in the ground

truth image. The number at corner indicates the corresponding correlation coefficient relative to the ground truth.

**High-resolution DOP and a non-binary object.**

Unlike low pixel numbers (~ a thousand) of a SPAD array, conventional cameras such as CMOS or CCD can easily provide megapixels. However, we probably do not need such high-resolution cameras due to the smooth distribution of diffusion photons at the observation plane. We attempt to explore the DOP performance with higher-resolution images and challenge it with non-binary objects (Fig. 5a). We run DOP with 3x3 illumination points to reconstruct the object in 2 scenarios, unknown and known diffusion characteristics, as presented in Fig. 5b and 5c, respectively. Two different resolutions (32x32 and 64x64) of images are fed into DOP algorithm. The object's dim region is revealed in all cases, while diffusion characterization allows reconstruction with a slightly sharper image. Using higher-resolution images results in more uniform reconstruction and a more obvious dim region. The reconstruction quality is significantly improved from 32x32 to 64x64 resolution; however, higher resolutions (beyond 64x64) do not significantly enhance image quality further while substantially increasing the computational cost, especially when there is no diffusion calibration.

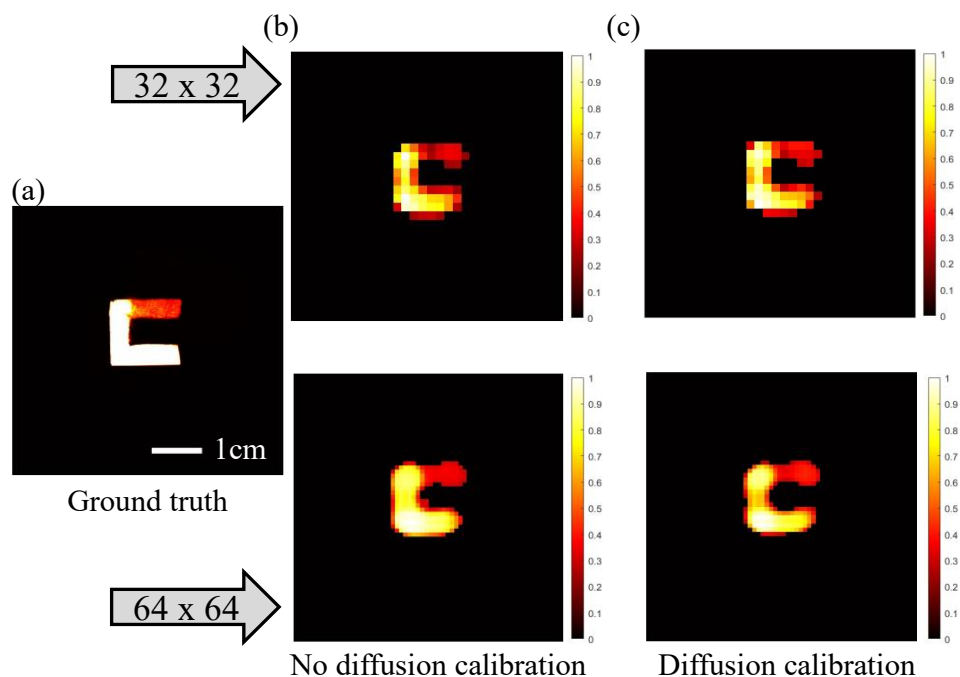

**Figure 5. DOP reconstruction results with higher resolution and a grayscale object. a)** Ground truth. DOP reconstruction results obtained with **b)** unknown and **c)** known diffusion characteristics. Two different resolutions are used for reconstruction 32x32 (top row) and 64x64 (bottom row). All figures have the same scale bar as presented in the ground truth image.

Resolving small features of embedded objects within thick diffusive media is extremely challenging due to a broad distribution of diffused photons. By exploiting the extra information of overlapping illuminations, we expect more advancements in our DOP technologies. We prepared 3 different samples where the square letter "C" has varying gaps of 5 mm, 2.5 mm, and 1 mm between its two arms (Fig. 6a). We conducted experiments with 3x3 illumination points and ran the DOP algorithm with and without calibration of the media's diffusion characteristics. Without diffusion calibration, the DOP technology remains effective, successfully resolving objects with feature sizes as small as 2.5 mm (Fig. 6b). However, if diffusion characteristics are properly measured and accounted for, our DOP method achieves significantly enhanced image sharpness, resolving features down to a 1 mm gap (Fig. 6c). These results underscore the substantial advantage of our DOP approach, significantly surpassing typical resolution limits of existing DOT approaches while remaining the simplest experimental setup.

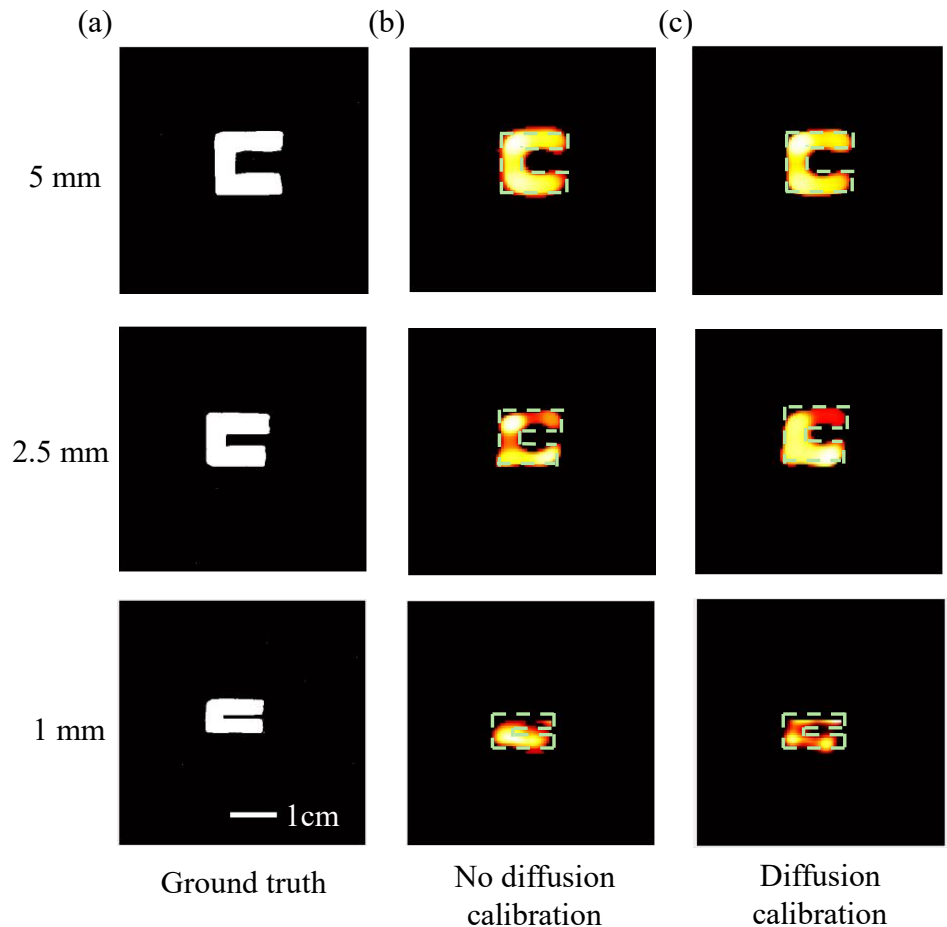

**Figure 6. DOP reconstruction images with objects of different feature sizes. a)** Three objects of square C with different gaps (5, 2.5 and 1mm). DOP reconstruction images under two different scenarios: **b)** uncharacterized and **c)** characterized diffusion properties. To resolve smaller feature of the object, we increased the resolution of images to 96x96.

# Conclusions and Discussions

Our research introduces an effective DOP technique combining a simple experimental setup with an efficient computational algorithm. Experimental and simulation results show minimal contribution of temporal information in the current state-of-the-art ToF-DOT due to very high correlation among the temporal frames. While the complex experimental setup and limitations of SPAD camera deteriorate the image reconstruction quality of ToF-DOT. Our approach only uses a conventional camera to capture multiple images corresponding to multiple illumination patterns. The fundamental advantages of our DOP approach derive from the utilization of spatially overlapping illuminations with minimal correlation, providing additional spatial information for accurate image reconstruction. Our novel technique demonstrates superior performance, achieving a high resolution down to 1 mm and producing high-quality reconstruction images for both binary and grayscale objects embedded within media thicker than 100 transport mean free paths. The reconstruction quality depends on multiple factors, some of which are entangled to each other such as the diffusion calibration, the optical diffusion properties and illumination conditions. Carefully calibrated diffusion properties allow higher reconstruction quality, but DOP still performs well without any calibration. High material diffusion coefficients or thicker media make a broad photon distribution (a broad PSF), which might not allow us to achieve overlapping illuminations with minimal correlation, especially on the small features of an embedded object. This poses a practical limit on the achievable resolution and underscores the importance of optimizing illumination strategies to address these inherent limitations. Our results present the significant potential of DOP for diverse applications, opening promising new avenues in medical imaging, non-destructive inspection, and deep-sea exploration, among others.

# Supporting Information

Supporting information is available from the link (will be inserted later).

# Acknowledgements

The authors especially thank Dr. Chau Nguyen at The University of Siegen, Germany for great discussions and suggestion. The authors would like to thank the Ministry of Education Singapore (MOE) for financial support under its grants RG140/23, MOE2016-T3-1-006(S), the National Research Foundation, Singapore (NRF) under its grant NRF-CRP29-2022-0003, the Science and Engineering Research Board, India (SERB): MTR/2021/000841, the Indo-French Centre for the Promotion of Advanced Research (CEFIPRA): IFC/SARI/2022/7143, and the Indian Institute of Technology Goa (IIT Goa) for their financial support.

# Conflict of Interest

The authors declare no conflict of interest.

# Author Contributions

C.D. and S.K.S. initiated the idea and supervised the research. C.D., M.H, and S.K.S, designed the experiments. M.H performed all the simulations, experiments, and analyzed the data. M.H fabricated the diffusive sample with the help of C.X. C.D and M.H wrote the manuscript with S.K.S.'s contributions. All authors discussed, analyzed, took responsibility for the results and revised the manuscript.

# Data Availability

The data and Matlab code that support the findings of this study are available from the corresponding author upon reasonable request.